\documentclass[a4paper,twocolumn,11pt,accepted=2022-01-22]{quantumarticle}
\pdfoutput=1
\usepackage[utf8]{inputenc}
\usepackage[english]{babel}
\usepackage[T1]{fontenc}
\usepackage{amsmath}
\usepackage{amssymb}
\usepackage{hyperref}
\usepackage{tikz}
\usepackage[numbers,sort&compress]{natbib}
\usepackage{graphicx}
\usepackage{caption, subcaption}
\begin{document}

\title{Minimal orthonormal bases for pure quantum state estimation}

\author{Leonardo Zambrano} 
\email{leonardo.zambrano@icfo.eu}
\affiliation{ICFO - Institut de Ciencies Fotoniques, The Barcelona Institute of Science and Technology, 08860 Castelldefels, Barcelona, Spain}
\orcid{0000-0001-7070-1433}

\author{Luciano Pereira}
\affiliation{Instituto de F\'{\i}sica Fundamental IFF-CSIC, Calle Serrano 113b, Madrid 28006, Spain}
\orcid{0000-0003-1183-2382}

\author{Aldo Delgado}
\affiliation{Instituto Milenio de Investigaci\'on en \'Optica y Departamento de F\'isica, Facultad de Ciencias F\'isicas y Matem\'aticas, Universidad de Concepci\'on, Casilla 160-C, Concepci\'on, Chile}
\orcid{0000-0002-8968-5733}

\maketitle

\begin{abstract}
    We present an analytical method to estimate pure quantum states using a minimum of three measurement bases in any finite-dimensional Hilbert space. This is optimal as two bases are insufficient to construct an informationally complete positive operator-valued measurement (IC-POVM) for pure states. We demonstrate our method using a binary tree structure, providing an algorithmic path for implementation. The performance of the method is evaluated through numerical simulations, showcasing its effectiveness for quantum state estimation. 
\end{abstract}
\section{Introduction}	
Quantum state estimation \cite{Paris2004}, the process of experimentally determining the complete description of a quantum system, is essential for numerous applications, ranging from quantum information processing to quantum simulation.  In a $d$-dimensional quantum system, states can be described by positive semi-definite complex matrices with unit trace. Therefore, quantum state estimation requires knowledge of the expectation values of at least $d^2-1$ linearly independent Hermitian operators. The traditional method for estimating these expectation values is by measuring the $d^2-1$ generalized Gell-Mann matrices \cite{SQT1,SQT2}. However, this approach demands significant experimental resources and time when $d$ is large. An alternative is to measure $d+1$ mutually unbiased bases \cite{MUB2,MUB3,MUB5,MUB6,MUB7}. While this set offers better scaling, it is still linear in $d$, and it remains unknown if mutually unbiased bases exist in arbitrary dimensions. The minimal choice for estimation is to measure just one informationally complete, positive operator-valued measurement (IC-POVM) \cite{SIC3, SIC2,SIC4,SIC5,SIC6}. However, implementing this measurement is challenging as it requires coupling the system with an ancilla and performing a projective measurement on the complete system.

Fortunately, if the state is known to be pure, the complexity of the state estimation problem can be significantly reduced \cite{Eisert2020, Chen2013, Stefano2019}. In this case, we require the measurements to form a pure-state informational complete (PSIC)-POVM, which is a POVM that contains at least $2d$ elements and characterizes any pure state but a set that is dense only on a set of measure zero \cite{SIC2}. It has been demonstrated that by combining the computational basis with two unambiguously discriminating POVMs, which yield a total of $3(d-1)$ independent measurement outcomes, any pure state can be reconstructed \cite{Ha2018}. Additionally, it has been shown that using two or three measurements over a system plus an ancilla is sufficient to determine a pure state, except for a limited set of ambiguities \cite{Wang2021}. 

When the problem is restricted to measurements on orthogonal bases, previous research has shown that at least four bases are required to determine any pure state in $d = 3$ and $d \ge 5$, while for $d=4$ it is unknown if three or four bases are enough \cite{Carmeli2015,Sun2020}. It has been shown that a pure state of a spin $s$ ($d=2s+1$) can be determined by measurements of spin components along three different axes, that is, with three measurement bases, up to a null-measure set \cite{Jean_Pierre_Amiet_1999}. This proposal does not provide an analytical relationship between the experimentally acquired data and the probability amplitudes of the unknown pure states, that is, it is not constructive. The absence of a simple estimator forces the use of statistical inference methods, such as maximum likelihood estimation \cite{Shang2017}. This method is mathematically formulated as a many-variable optimization problem and becomes infeasible in higher dimensions due to the excessive computational cost. This unwanted feature can be eliminated by increasing the number of measurement bases from 3 to 5 \cite{5B,5BFIJAS, CI5BB}, which allows to estimate any $d$-dimensional pure quantum state with post-processing based on analytic methods \cite{5B,5BFIJAS, CI5BB}. This proposal is adaptive, so measurements on the canonical basis are used to adapt the measurements in the following four bases. It has been shown that it is possible to reduce the number of bases to 3 \cite{3B}, although it still requires adaptability. In this case, measurements on two bases lead to a number of estimates that increase exponentially with the dimension. Measurements on a third basis single out an estimate by selecting the estimate with the highest value of the likelihood. The process is completely analytical, and it does not require maximization of likelihood but rather evaluation. However, in higher dimensions, likelihood provides a flat landscape, making it increasingly difficult to identify the most compatible estimate with third-base measurements. This proposal has recently been adapted to the case of pure multi-qubit quantum states \cite{pereira2021}, where $mn+1$ separable bases ($m\ge2$) estimate all pure states, except a null measure set, without the use of statistical inference methods. Adaptive compressed sensing techniques \cite{Ahn,Ahn2019} have also been successfully applied to estimate pure states. 

In this article, we present a method to estimate almost any pure quantum state in any finite-dimensional Hilbert space. The method uses three measurement bases. This number is optimal, as two bases are insufficient to construct a pure-state IC-POVM \cite{SIC2}. To prove the method, we first show how to estimate the relative phases of a quantum state given the measurement of a set of projectors. Then, we propose a protocol that reconstructs all the amplitudes and phases of the state using the previous result. We show that the required measurements form at least three bases and provide an algorithm based on a binary tree structure that outputs the states of the three bases. Our approach is constructive since the probability amplitudes of the unknown state can be obtained by analytically solving a set of $2\times2$ linear problems. In this way, resorting to computationally expensive statistical inference methods is unnecessary. In particular, the total time required to obtain an estimate is given by the time to evaluate the analytical solution of a  $2\times2$ linear problem multiplied by the total number of linear problems, which is a linear function of $d$. Therefore, the total time scales as
a linear function of the dimension $d$. We also show that our method is not limited to three bases, it leads to the construction of an arbitrary number of bases that allow estimation of the unknown state. Additionally, we assess the performance of the method through numerical simulations. In particular, we show that our proposal leads to a good estimation accuracy and that using a higher number of bases is preferable over increasing the ensemble size. Finally, we show that the proposed method can estimate pure states affected by white noise.

\section{Method}

This section introduces our tomographic method to estimate pure state with three measurement bases. The demonstration is divided into three sections. First, we demonstrate how to estimate the relative phase of a state that lies between two subspaces of the Hilbert space through measurements on two projectors. We then use this result to show that a binary tree protocol allows us to estimate all the relative phases necessary to reconstruct the pure state. Finally, we show that the measurements required for the protocol can be arranged in three measurement bases.

\subsection{Relative phase estimation}
Let us consider a Hilbert space $\mathcal{H}_\gamma$ of dimension $d_\gamma$ such that it is the direct sum of two orthogonal subspaces  $\mathcal{H}_\alpha$ and $\mathcal{H}_{\beta}$ of dimension $d_\alpha$ and $d_{\beta}$, respectively, that is, $\mathcal{H}_\gamma = \mathcal{H}_\alpha \oplus \mathcal{H}_{\beta}$. An arbitrary vector in  ${\cal H}_{\gamma}$ can be written as 
\begin{align}
	|\tilde{\psi}_{\gamma} \rangle = | \tilde{\psi}_{\alpha} \rangle + e^{i \varphi} |\tilde{\psi}_{\beta} \rangle, \label{psi_n} 
\end{align}
with $| \tilde{\psi}_{\alpha} \rangle$ in  ${\cal H}_{\alpha}$, $| \tilde{\psi}_{\beta} \rangle$ in ${\cal H}_{\beta}$ and $\varphi\in[0, 2\pi)$. We are interested in an equation for the relative phase $\varphi$ assuming that we know $| \tilde{\psi}_{\alpha} \rangle$ and $| \tilde{\psi}_{\beta} \rangle$. In order to do so, we define a set of $N$ projectors $\{ |\gamma_j \rangle \langle \gamma_j| \}_{j=1}^N$ such that $|\gamma_j \rangle = |\alpha_j\rangle + | \beta_j \rangle$, $|\alpha_j \rangle \in  {\cal H}_{\alpha}$ and $|\beta_j \rangle \in  {\cal H}_{\beta}$. Then, 
\begin{align}
	| \langle \gamma_j | \tilde{\psi}_{\gamma} \rangle |^2  = & |\langle \alpha_j |\tilde{\psi}_\alpha \rangle|^2 + |\langle \beta_j | \tilde{\psi}_\beta \rangle|^2 \nonumber \\
    &+ 2  \operatorname{Re} \left[e^{i \varphi} \langle \tilde{\psi}_\alpha | \alpha_j \rangle  \langle \beta_j | \tilde{\psi}_\beta \rangle \right]. \label{square_abs_dot} 
\end{align}
We now assume that the vector $| \tilde{\psi}_\gamma \rangle$ in Eq.~\eqref{psi_n} is part of a quantum state $| \psi \rangle$ that belongs to a Hilbert space $\mathcal{H} = \mathcal{H}_\gamma \oplus \mathcal{H}_\gamma^\perp$ of dimension $d\geq d_\gamma$. Measuring the system with a POVM that contains the projectors $\{ |\gamma_j \rangle \langle \gamma_j| \}_{j=1}^N$ we can estimate a set of probabilities $\{p_{j} \}_{j=1}^N$ such that $p_{j}=| \langle \gamma_j |\psi \rangle |^2 = | \langle \gamma_j | \tilde{\psi}_{\gamma} \rangle |^2$. Then, we define the quantities
\begin{align}
    \tilde{p}_j \equiv  \frac{| \langle \gamma_j | \tilde{\psi}_{\gamma} \rangle |^2 - |\langle \alpha_j | \tilde{\psi}_\alpha \rangle|^2 - |\langle \beta_j | \tilde{\psi}_\beta \rangle|^2}{2} \label{p_tilde} 
\end{align}
and
\begin{align}
    \Gamma_j  \equiv      \langle \tilde{\psi}_\alpha | \alpha_j \rangle  \langle \beta_j | \tilde{\psi}_\beta \rangle,\label{N}
\end{align}
which are known. Substituting Eqs.~\eqref{p_tilde} and \eqref{N} into Eq.~\eqref{square_abs_dot} we find a system of equations $\Gamma\vec{x}=\vec{y}$ for the relative phase $ \varphi$ between $|\tilde{\psi}_\alpha \rangle$ and $|\tilde{\psi}_\beta \rangle$, which is given by
\begin{align}
    \begin{pmatrix}
         \operatorname{Re} \left[\Gamma_1 \right] & - \operatorname{Im} \left[\Gamma_1 \right] \\
         \operatorname{Re} \left[\Gamma_2 \right] & - \operatorname{Im} \left[\Gamma_2 \right] \\
        \vdots & \vdots \\
         \operatorname{Re} \left[\Gamma_N \right] & - \operatorname{Im} \left[\Gamma_N \right]
    \end{pmatrix}	
    \begin{pmatrix}
        \cos \varphi\\
        \sin \varphi
    \end{pmatrix}
    = 	
    \begin{pmatrix}
        \tilde{p}_1\\
        \tilde{p}_2\\
        \vdots\\
        \tilde{p}_N
    \end{pmatrix}.
    \label{sist_eqs}
\end{align}
We need at least two different equations for the system to have a unique solution, which, in the case $N>2$, can be found through the Moore-Penrose pseudo-inverse given by $\Gamma^+=(\Gamma^\top \Gamma)^{-1}\Gamma^\top$ when $\Gamma$ has linearly independent columns. The calculation of the pseudo-inverse $\Gamma^+$ is performed with good quality when it is well-conditioned. This can be estimated by the condition number $\rm{Cond(\Gamma)}=\sigma_{\max}(\Gamma)/\sigma_{min}(\Gamma)$, where $\sigma_{\max}(\Gamma)$ and $\sigma_{min}(\Gamma)$ are the maximum and minimum singular values of $\Gamma$, respectively. A condition number of $\Gamma$ much larger than 1 indicates a poor inversion quality. In the particular case of measuring just two independent projectors, the solution of the linear system of equations is analytical and given by
\begin{align}
    e^{i\varphi} = i \frac{ \tilde{p}_{2} \Gamma_1^* - \tilde{p}_{1} \Gamma_2}{\operatorname{Im} \left[\Gamma_1 \Gamma_2^* \right]},\label{exp_fase}
\end{align}
whenever $\operatorname{Im} \left[ \Gamma_1 \Gamma_2^*\right] \neq 0$. Since we know $| \tilde{\psi}_\alpha \rangle$, $| \tilde{\psi}_\beta \rangle$ and $	e^{i \varphi}$, we have completely characterized the vector $| \tilde{\psi}_{\gamma} \rangle$ in Eq.~\eqref{psi_n}. When $\operatorname{Im} \left[ \Gamma_1 \Gamma_2^*\right]=0$, the phase $\varphi$ cannot be obtained since the two columns in the matrix are linearly dependent. However, if we employ the identity $\cos^2(\varphi)+\sin^2(\varphi)=1$, we can determine this phase from a single equation $\operatorname{Re}\left[\Gamma_k\right]\cos(\varphi)-\operatorname{Im}\left[\Gamma_k\right]\sin(\varphi)=\tilde{p}_k$ except for two sign ambiguities, 
\begin{align}
    e^{i\varphi_{\pm}} = \frac{1}{\Gamma_k}\left( \tilde{p}_k \pm \sqrt{\tilde{p}_k-|\Gamma_k|^2} \right). 
\end{align}
When we choose the measurements randomly, the event $ \operatorname{Im} \left[ \Gamma_1 \Gamma_2^*\right] = 0$ has probability zero (see Appendix \ref{bad_states}). However, small values of this quantity could decrease the estimation quality when working with a finite number of shots (or ensemble size). This happens when the matrix has an eigenvalue close to zero, and then the two equations in the system are close to being linearly dependent. Adding another randomly generated projector $|\gamma_3 \rangle \langle \gamma_3|$ usually alleviates this problem since the new equation in the system is probably linear-independent of the others.

\subsection{Pure state tomography}
The previous procedure can be generalized to reconstruct pure states in any dimension $d$. For this, we use a complete, full binary tree structure with $2d-1$ nodes, which we describe in Appendix \ref{sec:binary_tree}. Consider an arbitrary pure quantum state 
\begin{equation}
	|\psi_{1}^{(d)} \rangle =\sum_{k=1}^{d} c_k e^{i \varphi_k} | k \rangle \label{psi_d} 
\end{equation}
in a $d$-dimensional Hilbert space $\mathcal{H}_{1}^{(d)}$. The coefficients entering in $|\psi_1^{(d)} \rangle$ are such that $0 \leq c_k \leq 1$  and $ 0 \leq \varphi_k < 2 \pi $. Super-indexes denote dimension, and sub-indexes denote nodes in the tree.

To start the quantum state estimation protocol, we decompose the Hilbert space $\mathcal{H}_1^{(d)}$ into $d$ direct sums of one-dimensional subspaces as $\mathcal{H}_1^{(d)} = \mathcal{H}_{d}^{(1)} \oplus \mathcal{H}_{d+1}^{(1)}\oplus ...\oplus \mathcal{H}_{2d-1}^{(1)}$. Each Hilbert space $\mathcal{H}_m^{(1)}$ contributes with a coefficient
\begin{align}
	| \psi_m^{(1)} \rangle &= c_{k} e^{i \varphi_{k} }|k \rangle  \label{psi_1}
\end{align}
of the state $|\psi^{(d)} \rangle$, with $k = m - d +1$. Measuring the canonical basis we obtain a set of probabilities $\{p_k\}_{k=1}^{d}$ which can be used to estimate all the coefficients $\{c_k\}_{k=1}^{d}$ in Eq.~\eqref{psi_1}. Then, to fully characterize the state, we need to find the values of the phases $\{\varphi_k\}_{k=1}^{d}$.\\ \begin{figure}[t]
	\centering
	\includegraphics[width=1\linewidth]{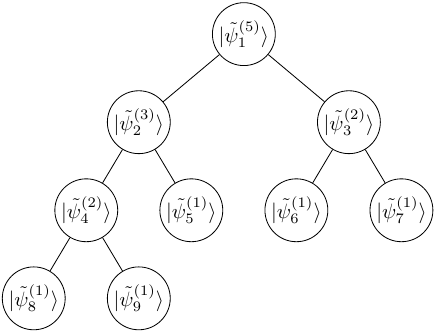}
	\caption{Tree structure needed to estimate a state in dimension $5$. Super-indices denote dimension, and sub-indices denote nodes in the tree.}
	\label{fig:tree_5}
\end{figure} 
We rely on a complete, full binary tree structure to find the phases $\varphi_{k}$. The tree has $2d-1$ nodes, and the leaves (nodes at the bottom of the tree) are $d$ vectors $| \tilde{\psi}_m^{(1)} \rangle = c_{k} |k \rangle $. Every internal node is a vector generated by the addition of its children and a relative phase, that is, 
\begin{align}
    |\tilde{\psi}_{m}^{(d_m)} \rangle = | \tilde{\psi}_{2m}^{(d_{2m})} \rangle + e^{i \varphi} |\tilde{\psi}_{2m+1}^{(d_{2m+1})} \rangle
\end{align}
in $\mathcal{H}_m^{(d_m)} = \mathcal{H}_{2m}^{(d_{2m})} \oplus \mathcal{H}_{2m+1}^{(d_{2m+1})}$, $d_{m}=d_{2m} + d_{2m+1}$. Let us assume that we know $| \tilde{\psi}_{2m}^{(d_{2m})} \rangle$ and $| \tilde{\psi}_{2m+1}^{(d_{2m+1})} \rangle$. Then, to obtain their parent node $|\tilde{\psi}_{m}^{(d_m)} \rangle$ we only need to find the phase $\varphi$. We do this by measuring projectors on $\mathcal{H}_m^{(d_m)}$ and using Eqs.~\eqref{p_tilde}, \eqref{N} and \eqref{sist_eqs}. Since we know all the leaves, we can apply this procedure recursively, starting from node $d-1$, and continuing with nodes $d-2$, $d-3$, and so on. We find the state $| \psi_1^{(d)} \rangle$ of the system when we reach the root of the tree (except for a global phase).

\begin{figure*}[t!]
	\begin{subfigure}[b]{0.32\textwidth}
		\includegraphics[width=\textwidth]{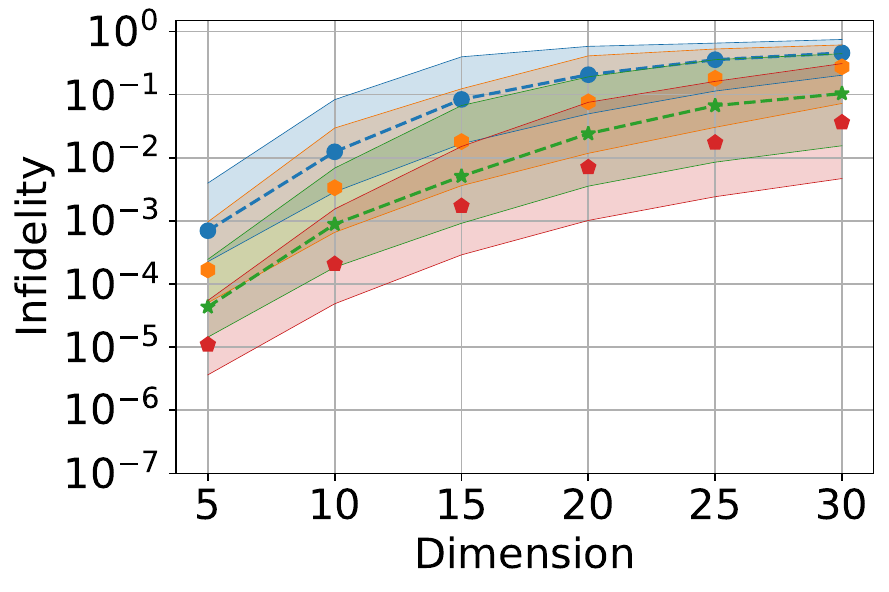}
		\caption{Random states estimated using $3$ bases.}
		\label{fig:Fig2-1}
	\end{subfigure}
	\begin{subfigure}[b]{0.32\textwidth}
		\includegraphics[width=\textwidth]{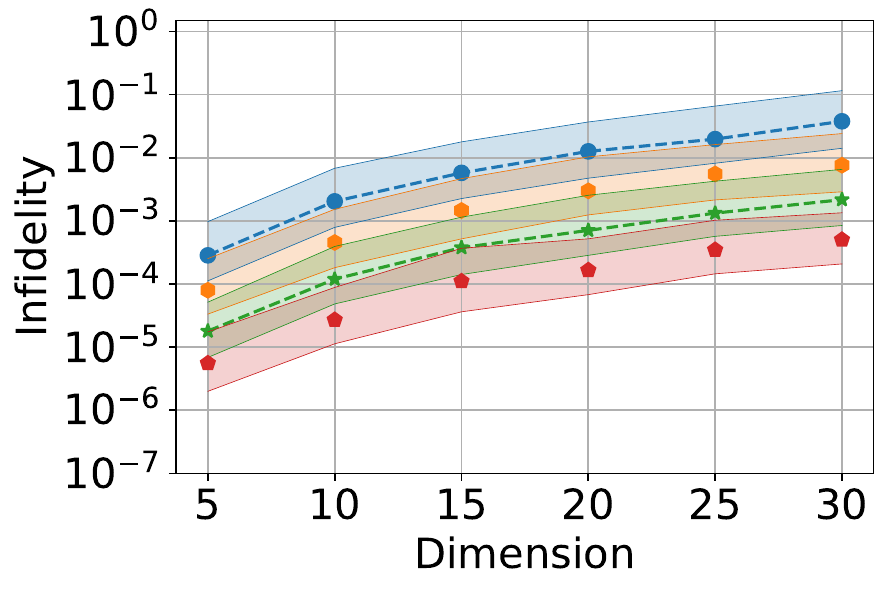}
		\caption{Random states estimated using $5$ bases.}
		\label{fig:Fig2-2}
	\end{subfigure}
	\centering
	\begin{subfigure}[b]{0.32\textwidth}
		\includegraphics[width=\textwidth]{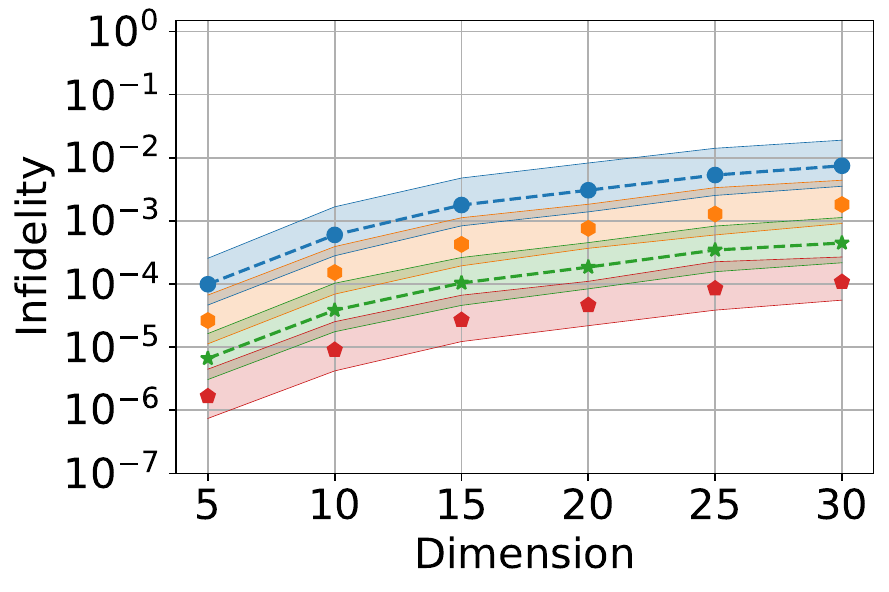}
		\caption{Random states estimated using $9$ bases.}
		\label{fig:Fig2-3}
	\end{subfigure}
	\caption{Median infidelity as a function of the dimension for $3$ (a), $5$ (b), and $9$ (c) bases with $2^{13}$ (blue), $2^{15}$ (orange), $2^{17}$ (green), and $2^{19}$ (red) shots. Shaded areas indicate the interquartile range.}
	\label{fig:Fig2}
\end{figure*}

Fig.~\ref{fig:tree_5} illustrates the tree for the case $d=5$. The leaves $m =9, 8, ..., 5$ contain vectors $| \tilde{\psi}_m^{(1)} \rangle = c_{m-4}|m-4\rangle$. Nodes $4$, $3$ and $2$ correspond to the states
\begin{align}
    |\tilde{\psi}_4^{(2)}\rangle &= c_{4} |4 \rangle + c_{5}e^{i( \varphi_{5} - \varphi_{4})}|5 \rangle, \\
    |\tilde{\psi}_3^{(2)}\rangle &= c_{2} |2 \rangle + c_{3}e^{i( \varphi_{3} - \varphi_{2})}|3 \rangle, \\
    |\tilde{\psi}_2^{(3)} \rangle &= c_{1}e^{i( \varphi_{1} - \varphi_{4})}|1 \rangle + c_{4} |4 \rangle  + c_{5} e^{i( \varphi_{5} - \varphi_{4})}|5 \rangle,
\end{align} 
and the root of the state 
\begin{equation}
|\tilde{\psi}_1^{(5)}\rangle = |\tilde{\psi}_2^{(3)}\rangle + e^{i( \varphi_{2} - \varphi_{4})} |\tilde{\psi}_3^{(2)}\rangle.
\end{equation}
This is the state of the system up to a global phase.

For the case of $n$-qubit systems, we have a perfect binary tree with $n$ levels. The vectors at the $n-1$ level are the same as the so-called $j=1$ substates of Ref.~\cite{pereira2021}. Since both methods take vectors and join them in pairs using direct sums, the internal nodes of the tree coincide with the substates. Furthermore, the systems of equations that must be solved at each step are the same, so our approach effectively generalizes the multi-qubit case to the qudit case.

\subsection{Measurement bases}

This section shows that the projective measurement required by the tomographic protocol can be arranged in a minimum number of three measurement bases. We first measure the canonical basis, giving us access to the amplitudes $\{ c_k\}$ of the pure state Eq.~\eqref{psi_d}. The other two measurement bases obtain the phases $\{\phi_k \}$. Since we have to measure at least $2$ projectors for each one of the $d-1$ internal nodes, the total number of projectors will be at least $2d-2$, which can be organized into $2$ measurement bases. A simple way to do this is to generate $d-1$ random vectors, one for each $\mathcal{H}^{(d_m)}$ with $d_m > 1$, in the tree, use the Gram-Schmidt method on the set, and complete with a last element to form a basis. Another way to obtain bases for the method involves a binary tree with the same structure as the one that we use to reconstruct $d$-dimensional states. Each leaf $m=d, ..., 2d-1$ now contains a vector in the computational basis
\begin{align}
	|s_m^{(1)} \rangle &=|m - d + 1 \rangle.
\end{align}
Each one of the $d-1$ internal nodes has associated two orthonormal vectors
\begin{align}
    \label{eq:r}
	|r_m^{(d_m)} \rangle &= a |s_{2m}^{(d_{2m})} \rangle + b e^{i \phi}|s_{2m+1}^{(d_{2m+1})} \rangle,\\
	|s_m^{(d_m)} \rangle &= b |s_{2m}^{(d_{2m})} \rangle - a e^{-i \phi}|s_{2m+1}^{(d_{2m+1})} \rangle,
\end{align}
with $a$, $b$ real, and $\phi$ a phase. These parameters are fixed for each basis and along all the nodes in the tree. The estimation method works for arbitrary choices of those parameters, but we note that setting $a=b=1/\sqrt{2}$ and $\phi$ uniformly distributed in $[0,\pi]$ delivers high fidelity for almost all pure states. Notice that all the $d-1$ vectors $|r_m^{(d_m)}\rangle$ are orthogonal to each other. Let us assume two arbitrary internal nodes $i, j$, with $i > j$ to prove this. If $i$ is not an ancestor of $j$, then $|r_i^{(d_i)}\rangle$ is orthogonal to $|r_j^{(d_j)}\rangle$, since, by construction, these vectors live in different orthogonal subspaces of $\mathcal{H}^{(d)}$. If $i$ is an ancestor of $j$, then $|r_i^{(d_i)}\rangle$ is of the form
\begin{align}
    |r_i^{(d_i)}\rangle = u |s_j^{(d_j)}\rangle + v| x \rangle,
\end{align}
where $|x\rangle$ is in ${\cal H}_{j}^{{(d_j) \perp}}$ and $u$ and $v$ are constants. This also means that $|r_i^{(d_i)}\rangle$ and $|r_j^{(d_j)}\rangle$ are orthogonal, as $\langle r_j^{(d_j)}|r_i^{(d_i)}\rangle = \langle r_j^{(d_j)}|s_j^{(d_j)}\rangle = 0$. Then, a basis for the protocol will be given by the set $\{|r_m^{(d_m)} \}_{m=1}^{d-1}$ plus a last element to complete the basis. We show an example of the bases for dimension $d=4$ in Appendix \ref{sec:bases}.

If we account for the canonical basis, the minimum number of bases needed to reconstruct arbitrary states in any dimension is $3$. For every internal node in the tree, we have to solve a system of equations in the form of Eq.~\eqref{sist_eqs}. With three bases, there might be some cases where some of these systems have no unique solution. If this happens, the maximum number of solutions that are consistent with the measurements is $2^{\lceil{n/2}\rceil-1}$. In Appendix \ref{bad_states}, we demonstrate that the probability of this event is negligible when we choose the bases at random. Despite this, if the bases are known, a state could be generated such that the estimation protocol fails. In Appendix \ref{sec:bases}, we show an example of this for dimension $d=4$, as well as the possible solutions of the estimation procedure. In such a case, we can always measure a fourth base to add more independent equations to the estimation protocol and obtain a unique estimator.

\section{Numerical simulations}
In order to test the performance of the method, we perform several numerical simulations. We randomly generate a set of $10^3$ Haar-uniform states $\{| \psi_i \rangle \}_{i=1}^{1000} $  for $d=5,10,15,20,25,30$. We apply the estimation protocol using $3$, $5$ and $9$ bases, whose measurement is simulated using $2^{13}$, $2^{15}$, $2^{17}$ and $2^{19}$ shots per basis. The bases in Eq.~\eqref{eq:r} are generated with the choice $a=b=1/\sqrt{2}$ and $\phi$ randomly, uniformly in $[0, 2\pi)$. As figure of merit for the accuracy of the estimation we use the infidelity  $I(| \psi_i \rangle, | \psi_i^{(\text{est})} \rangle) = 1  - |  \langle \psi_i  | \psi_i^{(\text{est})} \rangle|^2$ between the target states $| \psi_i \rangle$ and their estimates $| \psi_i^{(\text{est})} \rangle$. We expect this quantity to be close to zero.

Figure~\ref{fig:Fig2-1} shows the median infidelity for the protocol, obtained by measuring three bases as a function of the dimension. The different curves correspond to the different number of shots with $2^{13}$, $2^{15}$, $2^{17}$, and $2^{19}$, from top to bottom. Shaded regions indicate the interquartile range. Figs.~\ref{fig:Fig2-2} and \ref{fig:Fig2-3} show the results of the simulations for $5$ and $9$ bases, respectively. Figures~\ref{fig:Fig2-1},\ref{fig:Fig2-2} and \ref{fig:Fig2-3} indicate that, in general, the estimation protocol here proposed accurately estimates pure states. For a fixed number of bases or shots, the quality of the estimation decreases as the dimension increases. This is natural because as the dimension increases the number of parameters to be estimated also increases. For a fixed dimension, the estimation accuracy can be improved by increasing the number of shots. However, it is often more effective to increase the number of bases. For instance, using $3$ bases in dimension $30$, we achieve infidelities on the order of $10^{-2}$ using $3 \times 2^{19}$ total shots. Using $9$ bases with a total of $9 \times 2^{13}$ total shots gives better results for the same dimension.
        
\section{Discussion and Conclusions}

We have shown that it is possible to reconstruct pure quantum states in arbitrary dimensions using a minimum of three bases. These constitute an instance of PSIC-POVM, that is, our estimation procedure allows characterization of all qudit states up to a set of measure zero. Furthermore, our protocol is optimal in the sense that at least $2d-1$ independent POVM elements are required to form a PSIC-POVM \cite{SIC2}. As $2$ bases only contain $2d-2$ independent POVM elements, they are insufficient to determine a generic pure state, and thus, the measurement of a third basis is mandatory to generate a PSIC-POVM. Thus, our estimation protocol has the minimal number of bases required to form a PSIC-POVM. 

Our estimation protocol is formulated in terms of a binary tree structure. This provides a direct implementation of the protocol for any dimension and a recipe to build the required bases. The estimation procedure may fail for certain states. Even when these states form a null measure set, this problem can be overcome by adding at least one more basis. Adding more basis also increases the precision of the estimation for a finite number of shots, as our numerical simulations show.  

Recently, high-dimensional quantum systems have been renewed subject of interest for applications in quantum information, for instance, multi-core fibers \cite{carine2020multi,martinez2023certification}, orbital angular moment of light \cite{Willner2021, Rojas2021, Akatev2022}, and qudit-based quantum processors \cite{Lu2020phaseestimation,Chi2022,Ringbauer2022}. The characterization of this class of systems requires procedures with the least number of measurement results and efficient post-processing, such as the estimation procedure proposed here. Furthermore, as complementary results, we show that our procedure can also estimate the pure states affected by white noise in Appendix \ref{sec:white_noise}, and that its execution times scale efficiently with the dimension in Appendix \ref{appendix:times}. These properties make our protocol well-suited for use in high-dimensional experimental contexts.

A potential extension of this work is a better characterization of the states the method cannot reconstruct and explore subspaces where our PSIC-POVM would become an IC-POVM \cite{Rehacek2009}. It would be interesting to design an adaptive procedure that can detect, with just the measurement of the canonical basis, when a given state cannot be accurately estimated and then adjust the subsequent measurements accordingly. By doing so, we could potentially expand the range of pure quantum states that can be successfully reconstructed or even establish a "really" PSIC-POVM \cite{Finkelstein2004, Wang2018}. Furthermore, an extension to multi-qudit systems may be interesting for applications in quantum computing.

\begin{acknowledgments} 
LZ was supported by the Government of Spain (Severo Ochoa CEX2019-000910-S, TRANQI and European Union NextGenerationEU PRTR-C17.I1), Fundació Cellex, Fundació Mir-Puig and Generalitat de Catalunya (CERCA program). LP was supported by ANID-PFCHA/DOCTORADO-BECAS-CHILE/2019-772200275, the CSIC Interdisciplinary Thematic Platform (PTI+) on Quantum Technologies (PTI-QTEP+), and the Proyecto Sinérgico CAM 2020 Y2020/TCS-6545 (NanoQuCo-CM). AD was supported by ANID -- Millennium Science Initiative Program -- ICN17$_-$012 and FONDECYT Grants 1231940 and 1230586.
\end{acknowledgments}

\bibliographystyle{apsrev4-2}
\bibliography{bibliography}

\newpage    
\appendix

\onecolumngrid

\section{The set of states that cannot be estimated has measure zero}\label{bad_states}

Let $X$ be the set of states that the method cannot reconstruct. This set can be written as 
\begin{align}
    X = & \left\{\{|\psi \rangle \in \mathcal{H}_d : \text{exists $i$ even in the tree such that }  \operatorname{Im} \left[ \Gamma_1^{(i)} \Gamma_2^{(i)*}\right] = 0 \right\},
\end{align}
with $|\psi \rangle  = | \tilde{\psi}_{i} \rangle + e^{i \varphi} |\tilde{\psi}_{i+1} \rangle$ and $\Gamma_1^{(i)} \Gamma_2^{(i)*} = \langle \tilde{\psi}_i | \alpha_1 \rangle  \langle \beta_1 | \tilde{\psi}_{i+1} \rangle \langle \tilde{\psi}_{i+1} | \beta_2 \rangle  \langle \alpha_2 | \tilde{\psi}_i \rangle$. $X$ is the union of $d-1$ sets (one for each $i$ even in the tree) of the form
\begin{align}
    X_i = \left\{|\psi \rangle \in \mathcal{H}_d : \operatorname{Im} \left[ \Gamma_1^{(i)} \Gamma_2^{(i)*}\right] = 0 \right\}.
\end{align}
We must show that the probability of choosing a state in $X$ is zero,  $P(X) = 0$. Since $X$ is the finite union of the sets $X_i$, it is sufficient to show that $P(X_i) = 0$ for all $i$.  Let us fix $i$, $|\alpha_1 \rangle$, $|\alpha_2\rangle$, $|\beta_1\rangle$ and $|\beta_2\rangle$. The first case where $\operatorname{Im} \left[ \Gamma_1^{(i)} \Gamma_2^{(i)*}\right] = 0$ is simply when $\Gamma_1^{(i)}=0$ or $\Gamma_2^{(i)}=0$. The set of states that fulfill these conditions are
\begin{align}
    X_i = \left\{|\psi \rangle \in \mathcal{H}_d :\; |\psi_i\rangle\perp|\alpha_1\rangle,\; |\psi_i\rangle\perp|\alpha_2\rangle,\; |\psi_{i+1}\rangle\perp|\beta_1\rangle,\; \text{or} \;|\psi_{i+1}\rangle\perp|\beta_2\rangle \right\},
\end{align}
which is the union of subspaces of dimension $d-1$, and then, is of null-measure in $\mathcal{H}_d$. For the case where $\Gamma_1^{(i)}\neq0$ and $\Gamma_2^{(i)}\neq0$, we employ the fact that $\operatorname{Im}[xy] = 0$ if and only if $y = c x^*$, with $c$ a non-zero real number. Then, 
\begin{align}
    X_i = \left\{|\psi \rangle \in \mathcal{H}_d :\langle \tilde{\psi}_i | \left( | \alpha_1 \rangle  \langle \beta_1 | - c| \alpha_2 \rangle  \langle \beta_2 | \right)| \tilde{\psi}_{i+1} \rangle = 0 \right\}
\end{align}
Let be $A=| \alpha_1 \rangle  \langle \beta_1 | - c| \alpha_2 \rangle  \langle \beta_2 |$. Then, for a state to be unreconstructible, $| \tilde{\psi}_{i+1} \rangle$ must be in the kernel of $A$ or $| \tilde{\psi}_{i} \rangle$ must be in the kernel of $A^{\dagger}$. Since the rank of $A$ is $2$, these spaces are at most $d-2$ dimensional, and then, they are of null measure in the set of quantum states in $\mathcal{H}_d$.\\

\section{Binary tree}\label{sec:binary_tree}
\begin{figure}[h!]
    \centering
	\includegraphics[width=0.3\linewidth]{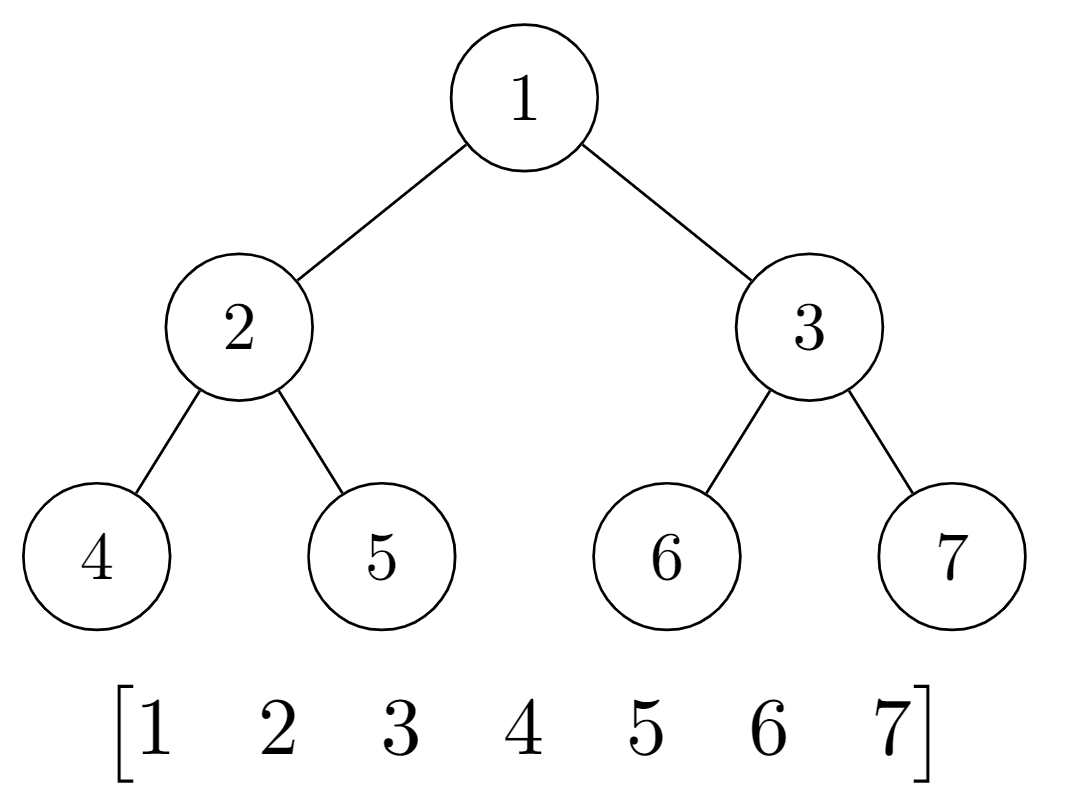}
	\caption{Full, complete, and perfect binary tree. The parent of the node at index $i$ is at index integer part of $i/2$.}
	\label{fig:fullbinarytree}
\end{figure}

A binary tree is a tree data structure in which each node has at most two children, referred to as the left and right child. It is full if every node has either 0 or 2 children, and it is complete if all the levels are completely filled except possibly the lowest one, which is filled from the left. Complete binary trees can be represented as an array, storing the root at position 1, the root's left child at position 2, and the root's right child at position 3. For the next level, the left child of the left child of the root is stored in position 4, and so on. See Figs.  \ref{fig:tree_5} and \ref{fig:fullbinarytree}. In the array, the left child of the node at index $i$ is at index $2i$, and the right child of the node at index $i$ is at index $2i+1$. Similarly, the parent of the node at index $i$ is at index integer part of $i/2$.\\
In a full binary tree 	
\begin{align}
	I = F - 1, 
\end{align}
where $F$ is the number of leaf nodes, and $I$ is the number of inner nodes. Finally, a perfect binary tree is a binary tree in which every inner node has exactly two children and all the leaf nodes are at the same level. Fig. \ref{fig:fullbinarytree} is also a  perfect binary tree. 

\section{Bases for dimension 4}\label{sec:bases}

Two possible bases for dimension $4$ are the following:
\begin{align}
B_1 = \begin{pmatrix}
\frac{1}{2} & 0 & \frac{1}{\sqrt{2}} & \frac{1}{2} \\
-\frac{1}{2} & 0 & \frac{1}{\sqrt{2}} & -\frac{1}{2} \\
 \frac{1}{2} & \frac{1}{\sqrt{2}} & 0 & -\frac{1}{2} \\
-\frac{1}{2} & \frac{1}{\sqrt{2}} & 0 & \frac{1}{2}
\end{pmatrix}, \qquad
B_2 = \begin{pmatrix}
\frac{1}{2} & 0 & \frac{1}{\sqrt{2}} &  \frac{1}{2} \\
-\frac{i}{2} & 0 & \frac{i}{\sqrt{2}} & -\frac{i}{2} \\
\frac{i}{2} & \frac{1}{\sqrt{2}} & 0 & -\frac{i}{2} \\
\frac{1}{2} & \frac{i}{\sqrt{2}} & 0 & -\frac{1}{2}
\end{pmatrix}
\end{align}  
Fig. \ref{fig:fullbinarytree} illustrates the binary tree corresponding to these bases. Each column is a vector $|\phi_i \rangle$ that defines a projector $| \phi_i \rangle \langle \phi_i |$, and the first $3$ columns of each matrix are associated with the first $3$ nodes of the tree.  The last column is to complete the basis.  \\
One state that cannot be reconstructed using these two bases, along with the canonical one, is
\begin{align}
    |\psi \rangle = \frac{1}{2} \left(|1\rangle + |2\rangle + |3\rangle + |4\rangle \right).
\end{align}
The vectors $|\tilde{\psi}_{2}^{(2)} \rangle = \frac{1}{2} (|1\rangle + |2\rangle)$ and $|\tilde{\psi}_{2}^{(3)} \rangle = \frac{1}{2} (|3\rangle + |4\rangle)$, corresponding to nodes $2$ and $3$ respectively, can be reconstructed without problem. For the root, we have
\begin{align}
        \Gamma_1  &= \frac{1}{4}\langle \tilde{\psi}_{2}^{(2)} | \left(|1 \rangle - | 2\rangle  \right) \left( \langle 3 |  - \langle 4 | \right)| \tilde{\psi}_{3}^{(2)} \rangle = 0,\\
        \Gamma_2  &= \frac{1}{4}\langle \tilde{\psi}_{2}^{(2)} | \left(|1 \rangle - i| 2\rangle  \right) \left( - i \langle 3 | + \langle 4 | \right)| \tilde{\psi}_{3}^{(2)} \rangle = -\frac{i}{8},
\end{align}
and $\tilde{p}_1 = \tilde{p}_2 = 0$. The first equation is $0=0$, which does not give any information about the quantum state. The system of equations is
\begin{align}
    \begin{pmatrix}
         0 & 0 \\
         0 & \frac{1}{8} 
    \end{pmatrix}	
    \begin{pmatrix}
        \cos \varphi\\
        \sin \varphi
    \end{pmatrix}
    = 	
    \begin{pmatrix}
        0\\
        0
    \end{pmatrix},
\end{align}
with solution $\varphi = 2\pi n$. From this, we have two possible reconstructed states,
\begin{align}
    |\psi_\pm \rangle = \frac{1}{2} \left(|1\rangle + |2\rangle \pm |3\rangle + |4\rangle \right).
\end{align}
Here, the problem appears because $\Gamma_1 = 0$. This depends on the value of the dots products between $|\tilde{\psi}_{2 (3)}^{(2)} \rangle$ and the first vectors of the basis $B_1$. Then, even the slightest change of the basis makes this problem solvable.
 
\section{States with white noise}\label{sec:white_noise}
A widely used model to characterize noise produced by the environment is white noise, which similarly affects all the eigenvalues of a density matrix. Arbitrary pure states affected by white noise are transformed into mixed states of the form
\begin{equation}
	\rho=(1-\lambda)|\psi\rangle\langle\psi|+\frac{\lambda}{d}\mathbb{I},
	\label{WhiteNoise}
\end{equation}
where $\mathbb{I}$ is the identity and $0 < \lambda \ll 1$ is the mixture parameter.\\
Let us now consider the impact of the white noise in the method. The measurement of a projector $|\gamma^{(2)}\rangle \langle \gamma^{(2)}|$, with $| \gamma^{(2)} \rangle  = a |k \rangle + b e^{i\phi} |k+1\rangle$, gives us
\begin{align}
\langle \gamma^{(2)} | \rho | \gamma^{(2)} \rangle  
=& (1 - \lambda) |c_{k}|^2 |a|^2 + \frac{\lambda}{d} |a|^2 + (1 - \lambda)|c_{k+1}|^2 |b|^2 + \frac{\lambda}{d} |b|^2 \nonumber \\
& + 2 (1 - \lambda)|a||b| c_{k}c_{k+1}  \operatorname{Re} \left[e^{-i \phi}  e^{i(\varphi_{k+1} - \varphi_{k})} \right].
\label{n_rho_n}
\end{align}
Measurements on the computational basis give us 
\begin{eqnarray}
	p_{k}^{(\rho)} &=& (1-\lambda)|c_{k}|^2 +\frac{\lambda}{d}. \label{probk}
\end{eqnarray}
We now define
\begin{align}
	\tilde{P}^{(\rho)} =& \frac{1}{|a||b|}\left(\langle \gamma^{(2)} | \rho | \gamma^{(2)} \rangle -  (1 - \lambda) |c_{k}|^2 |a|^2 - \frac{\lambda}{d} |a|^2 \right.  \left. -  (1 - \lambda)c_{k+1}^2 |b|^2 - \frac{\lambda}{d} |b|^2 \right)  \nonumber\\
		=& \frac{1}{|a||b|}\left(\langle \gamma^{(2)} | \rho | \gamma^{(2)} \rangle -   p_{k}^{(\rho)}  |a|^2 -  p_{k + 1}^{(\rho)} |b|^2 \right).
\end{align}
From the equation above, we see that 
$\tilde{P}^{(\rho)}$ can be expressed as a function of the measurements over the computational basis and $|\gamma^{(2)}\rangle$, besides the known parameters $a$ y $b$. With this, Eq. \eqref{n_rho_n} becomes
\begin{align}
	\tilde{P}^{(\rho)} = 2 (1 - \lambda)|a||b| c_{k}c_{k+1}  \operatorname{Re} \left[e^{-i \phi}  e^{i(\varphi_{k+1} - \varphi_{k})} \right].
\end{align}
Based on this, we measure a set of projectors $|\gamma_j^{(2)} \rangle \langle \gamma_j^{(2)}| $, and we get the following system of linear equations:
\begin{align}
	\begin{bmatrix}
		\cos \phi_1 & \sin \phi_1 \\
		\vdots & \vdots \\
		\cos \phi_f & \sin \phi_f 
	\end{bmatrix}
	\begin{bmatrix}
		2(1 - \lambda)c_{k} c_{k +1}\cos( \phi_{k + 1} - \phi_{k})\\
		2(1 - \lambda)c_{k} c_{k +1}\sin( \phi_{k + 1} - \phi_{k}) 
	\end{bmatrix} 
        =
	\begin{bmatrix}
		\tilde{P}^{(\rho)}_{1}\\
		\vdots \\
		\tilde{P}^{(\rho)}_{f}
	\end{bmatrix}\label{EqSyst1}.
\end{align} 
The solution for \eqref{EqSyst1} gives a value for
\begin{align}
	\Lambda^{(\rho)} =  2(1 - \lambda)c_{k } c_{k +1} e^{i( \phi_{k + 1} - \phi_{k})}, \label{lambdas}
\end{align}
Taking the absolute value of \eqref{lambdas} we obtain
\begin{align}
	|\Lambda^{(\rho)}|^2 =  4(1 - \lambda)^2|c_{k }|^2 |c_{k +1}|^2. \label{abs_lambdas}
\end{align}
Replacing  $|c_{k}|^2$ and $|c_{k+1}|^2$ from Eq. \eqref{probk} we get
\begin{align}
	| \Lambda^{(\rho)}|^2 = 4\left(p_{k}^{(\rho)}-\frac{\lambda}{d}\right)\left(p_{k+1}^{(\rho)}-\frac{\lambda}{d}\right),
\end{align}
which is a quadratic equation for $\lambda$ with solution
\begin{align}
	\lambda=\frac{d}{2}\left[ p_{k}^{(\rho) }+p_{k + 1}^{(\rho) } - \sqrt{(p_{k}^{(\rho) }-p_{k+1}^{(\rho) })^2+|\Lambda^{(\rho)}|^2}     \right], \label{EstimatedLambda}
\end{align}	
where we chose the solution with a minus sign since  $\lambda$ must be small. From this, we can estimate a value for the purity of the system. We can also correct the values of the coefficients $\{c_k\}_{k=1}^d$ using Eq. \eqref{probk}, and, through this, obtain a corrected estimation of $| \psi \rangle$.\\

\section{Benchmarking pure tomography methods with three bases}\label{appendix:times}

The problem of determining the minimal number of measurements for reconstructing all pure quantum states has been extensively investigated in the literature \cite{Eisert2020, Chen2013, Stefano2019,SIC2,Ha2018,Wang2021,Wang2021,Carmeli2015,Sun2020}. In particular, in Ref.\cite{Jean_Pierre_Amiet_1999} it had been demonstrated that all pure states can be determined up to a set of measure zero by measurements in three spin observables $O_j = \vec{s}_j\cdot\vec{\sigma}$, with $\{\vec{s}_j\}$ non-coplanar vectors and $\vec{\sigma}=(\sigma_x,\sigma_y,\sigma_z)$ a $d$-dimensional irreducible representation of the SU(2) group. Despite proving that a pure state can be determined with these three observables, it does not provide a constructive method to obtain the pure state from the estimated probabilities, forcing the usage of statistical inference methods, such as maximum likelihood estimation, to reconstruct the state. 

\begin{figure}[h!]
    \centering
    \includegraphics[width=\linewidth]{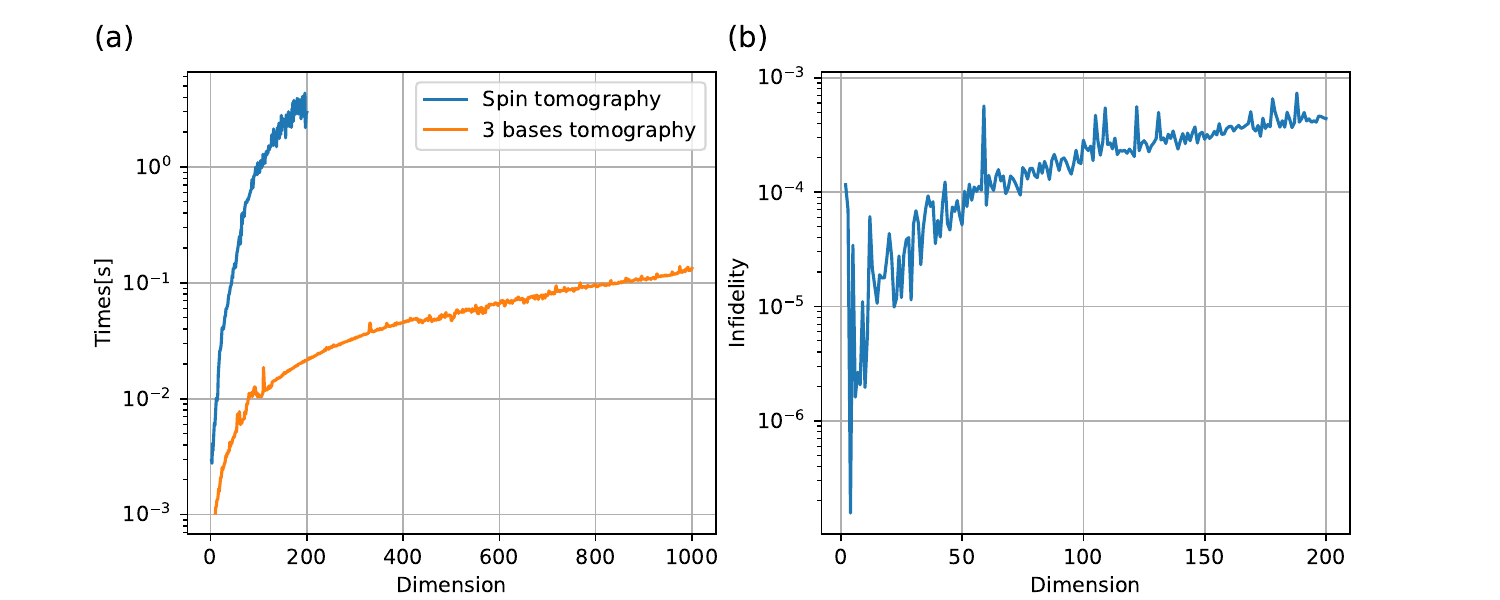}
    \caption{ a) Average execution time of the pure state tomography with our three-base protocol (blue) and the three spin bases with SF-MLE (orange). b) Average infidelity of the estimated pure states with the three spin bases with SF-MLE. The averages are calculated over 100 randomly generated pure states.}
    \label{fig:times}
\end{figure}

We perform numerical simulations to compare the proposal from \cite{Jean_Pierre_Amiet_1999} with our three-base estimation protocol. Superfast maximum likelihood estimation (SF-MLE) \cite{Shang2017} is used to estimate states from measurements on the three spin bases. We randomly generate a set of 100 pure states for each dimension and simulate their tomographic reconstruction with ideal probabilities, that is, with an infinite number of shots. Figure~\ref{fig:times}a shows the average execution time of both methods for a wide range of dimensions, and Fig.~\ref{fig:times}b displays the average infidelity between the actual state and its estimated counterpart in the context of the three spin bases approach. Notice that our method always gives perfect infidelity of zero, so we prefer not to put it in the figure. The figures show that our proposal outperforms the execution time of the tomography with three spin bases and SF-MLE, having a more favorable scaling with the dimension. Particularly, our method can reconstruct pure states of dimension 1000 in 0.33 seconds with perfect fidelity, while the three spin bases with SF-MLE reconstruct pure states of dimension 200 in 3 seconds and with a fidelity of 0.9995. Therefore, our method is faster and more accurate. 

\begin{figure}[h!]
    \centering
    \includegraphics[width=1\linewidth]{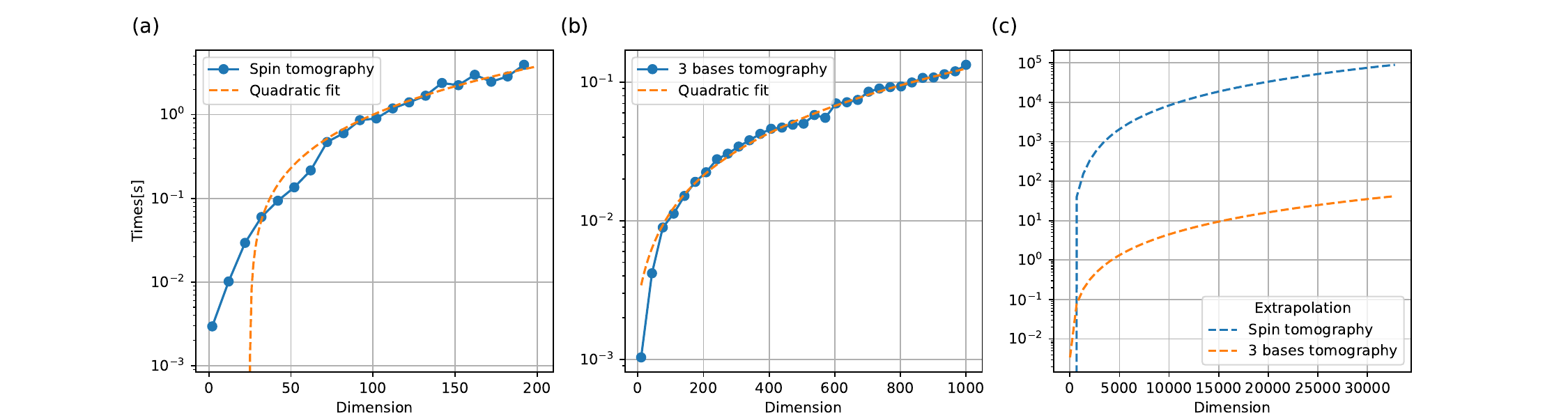}
    \caption{Quadratic fits of the execution time for a) the three-base protocol and the b) three spin bases with SF-MLE. For display purposes, we do not include the data of all dimensions, but the fits were done considering all of them. c) Extrapolated execution time until dimension $2^{15}$.}
    \label{fig:fits}
\end{figure}

Figure~\ref{fig:fits} depicts quadratic fits of the execution time for both methods, with an extrapolation of the execution time up to dimension $2^{15}$. The coefficients of the fits, detailed in Table \ref{tab:fits}, demonstrate a proximity to linear trends. The extrapolated time of our proposal for dimension $10^5$ is less than 10 seconds. In contrast, the extrapolated time for SF-MLE with three bases is approximately 3 hours, approximately $10^3$ times more than our method.

\begin{table}[h!]
    \centering
    \begin{tabular}{|c|c|c|c|}
    \hline
        Method & a & b & c \\
     \hline
      \hline
        Spin bases & $8.26389347\times10^{-5}$ & $2.99062971\times10^{-3}$ & $-1.25158952\times 10^{-1}$ \\
         \hline
        Three bases & $3.61914324\times10^{-8}$ & $8.62827934\times10^{-5}$ & $2.55736509\times 10^{-3}$\\
         \hline
    \end{tabular}
    \caption{Coefficients of the quadratic fit $ad^2+bd+c$ of Fig.~\ref{fig:fits}.}
    \label{tab:fits}
\end{table}

\end{document}